\documentclass[pra, a4paper, twocolumn, 10pt, showpacs]{revtex4}
\usepackage{bbm, amsmath, amssymb, amsthm, bm,textcomp, nicefrac}
\usepackage[T1]{fontenc}
\usepackage[latin9]{inputenc}
\usepackage{graphicx}
\usepackage{geometry}

\newcommand{\one}{\mathbbm{1}}

\newcommand{\ket}[1]{\left|{#1}\right\rangle}
\newcommand{\bra}[1]{\left\langle{#1}\right|}

\newcommand{\ketbrad}[1]{\left|{#1}\rangle\!\langle{#1}\right|}
\newcommand{\ketbra}[2]{\left|{#1}\rangle\!\langle{#2}\right|}

\newcommand{\be}{\begin{equation}}
\newcommand{\ee}{\end{equation}}
\newcommand{\eea}{\end{eqnarray}}
\newcommand{\bea}{\begin{eqnarray}}

\begin{document}

\title{Stable macroscopic quantum superpositions}

\author{F.\ Fr\"owis and W.\ D\"ur}

\affiliation{Institut f\"ur Theoretische Physik, Universit\"at
  Innsbruck, Technikerstr. 25, A-6020 Innsbruck,  Austria}
\date{\today}

\begin{abstract}
We study the stability of superpositions of macroscopically distinct quantum states under decoherence. We introduce a class of quantum states with entanglement features similar to Greenberger-Horne-Zeilinger (GHZ) states, but with an inherent stability against noise and decoherence. We show that in contrast to GHZ states, these so-called concatenated GHZ states remain multipartite entangled even for macroscopic numbers of particles and can be used for quantum metrology in noisy environments. We also propose a scalable experimental realization of these states using existing ion-trap set-ups.
\end{abstract}


\maketitle


Ever since quantum mechanics was established as a theory describing light and matter at a microscopic level, the logical implications and its possible extension to the macroscopic realm have been subject to vivid debates. Already in 1935 Schr\"odinger \cite{Schroedinger35} pointed out in his famous Schr\"odinger-cat gedanken experiment that quantum mechanics predicts the existence of superpositions of macroscopically distinct states, thereby highlighting the counterintuitive features of quantum mechanics at a macroscopic scale. While this was not meant as a serious experimental proposal, remarkable progress towards an experimental realization of such ``cat-states'' has been reported in the last decade (see e.g. \cite{Experiments,IonsBlatt}). The interest in these states is not only triggered by fundamental considerations, but also holds a more practical perspective due to possible applications in quantum metrology \cite{helstrom,Caves,braunstein,holland,Giovanetti}.

An archetypal ''cat-state'' is the multipartite Greenberger-Horne-Zeilinger (GHZ) state
\be
\ket{GHZ_{N}^{\pm} }= \frac{1}{\sqrt{2}}(\ket{0}^{\otimes N} \pm \ket{1}^{\otimes N}).
\ee
While such a state can be argued to constitute a macroscopic quantum superposition and allows for parameter estimation surpassing the standard quantum limit, it is also widely believed that the influence of noise and decoherence renders the observation of related interference effects at a macroscopic level very difficult.
It has indeed been pointed out with various arguments \cite{simon,Du02,distill,aolita,Huelga} that the GHZ state is unstable under decoherence in the sense that its quantum properties such as entanglement or quantum coherence as well as its applicability for parameter estimation vanish for macroscopic system size practically instantaneously, with a rate growing exponentially with the size of the system.

In this letter we introduce a class of quantum superposition states that shows similar features as the GHZ state, but remain multiparticle entangled even for macroscopic system sizes and in noisy environments. The basic idea is to encode the states $\ket{0}$ and $\ket{1}$ by small groups of qubits such that quantum properties can be conserved even under decoherence. This is similar to the protection of general quantum information using quantum error correction, however no active correction or feedback is required. We study the action of decoherence on such encoded superposition states, which we model by single-particle white noise. We however point out that our observations are not restricted to this particular noise model, but are a generic feature of these states.

We illustrate the stability of the states by calculating several quantities which are commonly seen as typical quantum properties and show that they are exponentially more stable under decoherence than for the standard GHZ state. These quantities include the trace norm of quantum coherence terms, entanglement measures such as the negativity or multipartite distillability, as well as the Fisher information. The latter allows us to conclude that the states in question can be used for parameter estimation surpassing the standard quantum limit even in a noisy environment. Finally, we propose a scalable experimental realization of these states using existing ion-trap set-ups.


\textit{Definitions and notation.---}Our setting is the Hilbert space of $N m$ two-level systems, where $N$ is the number of logical qubits, each built of $m$ physical qubits. The state of interest is then defined as
\begin{equation}
  \label{eq:3}
\ket{\phi_C} = \frac{1}{\sqrt{2}}(\ket{GHZ_m^{+}}^{\otimes N} + \ket{GHZ_m^{-}}^{\otimes N})
\end{equation}
and is called concatenated GHZ (C-GHZ), since it exhibit a GHZ-like superposition of a $N$-fold tensor product of GHZ states. Note that for $m=1$ it simplifies to the standard GHZ state up to a local basis change. We consider groups of size $m$ as blocks constituting one logical qubit, and are interested in (entanglement) properties of the states with respect to these blocks. That is, we consider each block as one logical system on which we allow for joint operations.

We study the stability of states under uncorrelated local (i.e. single qubit) decoherence processes,
 \begin{equation}
  \label{eq:1}
  \rho_{\phi_C} = {\cal E}\ketbrad{\phi_C} = {\cal E}_1{\cal E}_2 \ldots {\cal E}_N\ketbrad{\phi_C},
\end{equation}
where 
${\cal E}_i\rho = p \rho + \frac{1-p}{4} \sum_{j=0}^3 \sigma_j^{(i)} \rho \sigma_j^{(i)}$
corresponds to a single qubit depolarizing (white noise) channel acting on qubit $i$ with error parameter $p=\exp(-\kappa t)$, and $\sigma_j$ denote Pauli operators with $\sigma_0=\one$. Notice that this uncorrelated noise model is rotational invariant with respect to single qubit operations and guarantees that no preferred basis with increased stability exists. Basis dependent noise models or correlated noise models typically allow to identify preferred basis or even decoherence free subspaces, thereby offering the possibility to identify states that are stable under such a particular noise process. While this is certainly of interest and practical relevance in many physical situations, 
here we are not interested in such basis-dependent stabilization effects. We rather would like to identify states that are stable under {\em any} (local) noise process and not just for some special noise model \cite{note1}.


\textit{Stability of the coherence terms.---} We start by considering the decay of coherences of the state (\ref{eq:1}), $\rho_{O} =
{\cal E}\ketbra{GHZ_m^{+}}{GHZ_m^{-}}^{\otimes N}$. These off-diagonal elements distinguish the coherent superposition of two macroscopically distinct states from a classical mixture, and can therefore be seen a characterizing the ``quantumness'' of the state. For {\em any} state of the form  $|a\rangle^{\otimes N} + |b\rangle^{\otimes N}$, the trace norm $||A||_1={\rm tr}\sqrt{A^\dagger A}$ of the corresponding coherences $\ketbra{a}{b}^{\otimes N}$ always decreases under local white noise exponentially with $\alpha N$. However, for $\rho_O$  we find that $\alpha$ can be made arbitrary small by increasing $m$.

To calculate $||\rho_{O}||_1$, it is sufficient to consider one logical qubit $\rho_{O1}={\cal E}\ketbra{GHZ_m^{+}}{GHZ_m^{-}}$, as $||\rho_{O}||_1=||\rho_{O1}||_1^N$. A direct calculation yields $||\rho_{O1}||_1 = 1-2^{1-m}\sum_{k = \lceil m/2\rceil}^m\binom{m}{k}\left(1+p\right)^{m-k}\left(1-p\right)^{k}$. For weak noise we can bound this sum from below using Stirling's formulae and find
\begin{equation}
  \label{eq:4}
  ||\rho_{O}||_1\geq \left[ 1-\sqrt{\frac{2m}{\pi}} \bigg( 1+\frac{1}{11m} \bigg) \bigg( 1-p^2 \bigg) ^{m/2} \right]^N.
\end{equation}
For fixed noise parameter $p$, $||\rho_{O}||_1$ tends to one exponentially fast with increasing $m$. In fact, by letting $m$ itself scale {\em logarithmically} with $N$, $m = O(\log(N))$ we find that the exponential decay of $||\rho_{O}||_1$ can be effectively frozen as $ \lim_{N \to \infty} ||\rho_{O}||_1 \to 1$ in this case, i.e. in order to stabilize the off-diagonal elements the number of physical qubits per logical block has to grow only logarithmically with the number of logical qubits.


\textit{Entanglement properties.---} We now turn to multipartite entanglement properties of the state with respect to the blocks constituting the logical qubits. We investigate the distillability of the decohered states, as well as the decay of certain multipartite entanglement measures. A $N$-party mixed state $\rho$ is called $N$-party distillable entangled if one can generate by means of local operation and classical communication any $N$-party entangled pure state --in particular a GHZ state-- from many copies of $\rho$ \cite{Du99}. The state $\rho$ remains multipartite entangled in this sense. It has been shown for general noise models that the GHZ state looses its multipartite distillability faster with increasing $N$ \cite{distill}, leading to a lifetime of distillable entanglement that vanishes practically instantaneously for large system sizes. Similarly, for any fixed noise parameter there is a maximal number of blocks that can remain multiparty entangled. In contrast, we now show that the C-GHZ state remains $N$-party distillable even for macroscopic $N$ by providing an explicit distillation protocol.

First, every logical qubit of $\rho_{\phi_C}$ is projected into the two-dimensional subspace spanned by $\ket{0}^{\otimes m}$ and $\ket{1}^{\otimes m}$. The resulting state can be viewed as a $N$--qubit state consisting of the logical qubits $|0_L\rangle=|0\rangle^{\otimes m}$ and $|1_L\rangle=|1\rangle^{\otimes m}$. The state is up to a logical Hadamard operation \cite{note2} diagonal in the GHZ-basis, and its distillability can be obtained directly \cite{Du99}. A measurement of all but two of the logical qubits in the basis $\{\ket{0_L},\ket{1_L}\}$ results into a two-party state. The fidelity of $\rho$ with respect to the (logical) Bell-state $|\phi^+_L\rangle=(|0_L0_L\rangle+|1_L1_L\rangle)/\sqrt{2}$ is given by $F = \bra{\phi_L^{+}} \rho \ket{\phi_L^{+}} = \frac{1}{4}\left\{ 1+\frac{p^{2m}}{d^2}\left[1+(\frac{o}{d})^{N-2}\right]+(\frac{o}{d})^N \right\}$, where $d = [(1+p)^m+(1-p)^m]/2^m$ and $o = [(1+p)^m-(1-p)^m]/2^m$. As long as $F > 1/2$, this two qubit state is distillable, from which multipartite distillability follows as the results holds for any pair of logical qubits \cite{Du99}. One observes that by increasing $m$, the range of distillability can be exponentially extended. If we consider e.g. $p=0.9$, the size of the GHZ state is limited to $N=53$ \cite{distill}. However, already a block size of $m=10$ allows for entanglement up to $N=10^{12}$ in the C-GHZ state,
i.e. the state remains distillable for a non-vanishing time even for this macroscopic system size. If we consider again blocks growing logarithmically with $N$, $m=O(\log(N))$, we can approximate for small noise $\left(\frac{o}{d}\right)^N \approx 1 - 2\left[e(1-p)(1+p)^{-1}\right]^{m}$ and we find $F>1/2$ for any finite $N$. Notice that also for the GHZ state, grouping several qubits to form a block allows one to increase the lifetime of distillable entanglement with respect to these blocks \cite{distill}, however a {\em linear} growth of block size, $m=O(N)$  with respect to the number of blocks $N$ is required to stabilize the state in contrast to the {\em logarithmic} growth, $m=O(\log N)$ for the C-GHZ state.

Secondly, we consider the decay rate of entanglement, where we use the negativity with respect to bipartite splits as measure \cite{neg}. We concentrate on the bipartition one (logical) qubit vs. rest, as this is the most fragile bipartition for the GHZ state \cite{simon,aolita} and the C-GHZ state. Entanglement with respect to other bipartitions decays slower and remains non-zero for longer times. In \cite{aolita} it has been shown that the negativity of the GHZ state shows an exponential decay, both with respect to time and system size $N$, leading to negligible values of entanglement for large $N$. Here we show that the decay rate with respect to $N$ can be made arbitrarily small by increasing the block size $m$.

\begin{figure}[ht]
  \begin{picture}(210,95)
\put(-15,-5){\includegraphics[width=120pt]{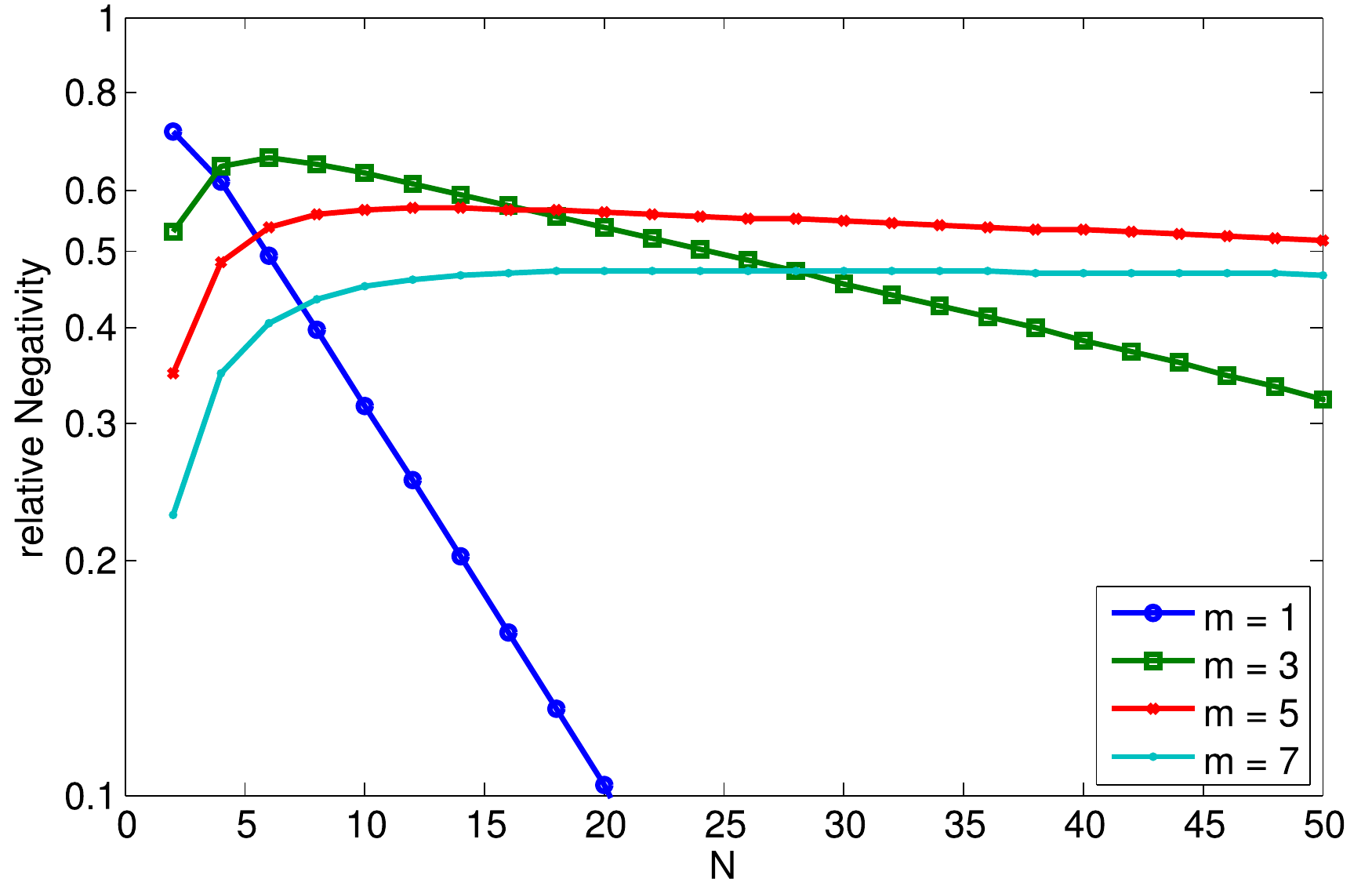}}
\put(110,-9){\includegraphics[width=120pt]{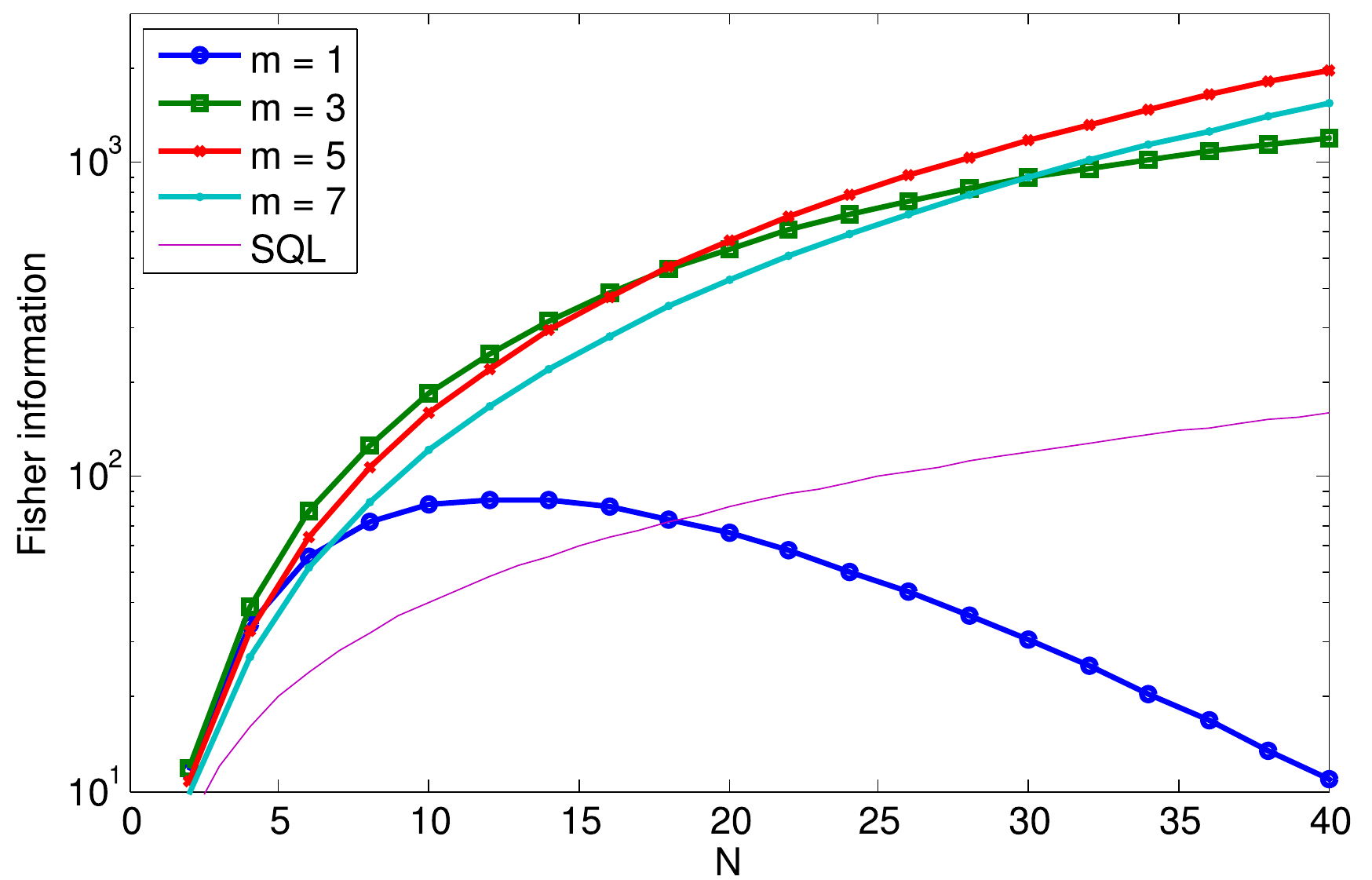}}
\put(45,103){(a)}
\put(171,103){(b)}
\end{picture}
\caption[]{\label{fig:Neg}\label{fig:Fish} (a) Relative negativity ${\cal N}(\rho_{\phi_C})/{\cal N}_0$ for the bipartition one logical qubit vs. rest for fixed noise parameter $p=0.9$ as a function of system size $N$, where the negativity of the initial state ${\cal N}_0=0.5$ for all $m$ and $N$. Notice the logarithmic scale. (b) Fisher information ${\cal F}_A(\rho_{\phi_C})$ for fixed noise parameter $p=0.9$ and $A = \sum_k (\sigma_x^{\otimes m})^k,$ as function of system size $N$ for different block sizes $m$. For comparison, the region between the standard quantum limit (SQL) and the Heisenberg limit is shaded.}
\end{figure}

The negativity of $\rho_{\phi_C}$ is defined as ${\cal N}(\rho_{\phi_C}) = \frac{1}{2}\left( ||\rho_{\phi_C}^{T_1}||_1-\mathrm{Tr}(\rho_{\phi_C}^{T_1}) \right)$, where partial transpositions is meant with respect one logical qubit. By applying a logical Hadamard rotation \cite{note2} to each logical qubit, the resulting operator becomes block diagonal and can be directly diagonalized. We obtain an analytic formula involving a sum of exponentially many eigenvalues. We find that the number of negative eigenvalues growth like a multinomial coefficient with respect to $N,m$, while the modulus decreases exponentially. For system sizes up to $m=7$ and $N=50$, we can evaluate this expression numerically, see Fig. \ref{fig:Neg}a. We observe also for large $N$ a significant entanglement even under the influence of decoherence, where the negativity decays exponentially with $N$. However, the decay rate {\em decreases} with growing $m$. In fact by fitting the tail to an exponential function $a\exp(-\gamma N)$, we obtain a rate $\gamma$ that decays itself exponentially with $m$, leading to non-vanishing negativity even for large $N$.



\textit{Fisher information.---}We now turn to quantum metrology \cite{helstrom,Caves,braunstein,holland,Giovanetti} and show that the C-GHZ states can be used for parameter estimation with a sensitivity above the standard quantum limit also in the presence of decoherence. Here a given quantum state $\rho$ evolves under $U=\exp(-i\theta A)$ and is then measured to estimate the unknown parameter $\theta$, where $A$ is a local generating operator. The confidence interval for parameter estimation can be bounded from above by the Cram\'er-Rao inequality $\delta \theta \geq \frac{1}{\sqrt{n {\cal F}_A(\rho)}}$ \cite{braunstein}, where ${\cal F}_{A}(\rho)$ is the so called Fisher information and $n$ denotes the number of repetitions of the experiment. For mixed states with spectral decomposition $\rho = \sum_k\lambda_k\ketbrad{k}$ the Fisher information reads \cite{braunstein}
\begin{equation}
{\cal F}_{A}(\rho) = 2\sum_{j,k} (\lambda_k-\lambda_j)^2/(\lambda_k+\lambda_j)|\bra{k}A\ket{j}|^{2}\label{eq:6}.
\end{equation}
If the Fisher information is of order $N$ --which corresponds to the standard quantum limit and is the upper limit for separable states \cite{Giovanetti}-- no gain compared to classical protocols can be achieved. The GHZ state --as well as other entangled states-- are however capable to reach the so-called Heisenberg limit, where ${\cal F}_A$ is $O(N^2)$ for the (optimal) choice $A=\sum_{k=1}^N \sigma_z^k$ \cite{holland}, thereby leading to a quadratic improvement in the achievable precision. Under general decoherence processes the GHZ state however quickly looses its advantages and especially for large $N$ there is no substantial gain as compared to uncorrelated particles \cite{Huelga}.

For the C-GHZ state we show that $\rho_{\phi_C}$ remains for sufficiently large $m$ above the standard quantum limit even in the presence of decoherence, and therefore stays useful for parameter estimation. We consider local generators consisting of sums of local interactions $A_C = \sum_{k=1}^N (\sigma_x^{\otimes m})^k$, where local is again understood in the sense of logical qubits. This means that for a given $m$, interactions involve blocks of $m$ physical qubits  $\sigma_x^{\otimes m}$.
For the pure state $\ket{\phi_C}$ this yields ${\cal F}_A(\ket{\phi_C}) =4N^2$ as for the GHZ state. We now investigate whether the decohered state $\rho_{\phi_C}$ is after some fixed time still useful for parameter estimation by calculating ${\cal F}_A(\rho_{\phi_C})$. After a logical Hadamard operation on each block \cite{note2}, the corresponding eigenvalues and eigenvectors can again be directly obtained due to the block-diagonal structure. We find that even for large $N$ the Fisher information stays well above the standard quantum limit, despite the exponentially decay with respect to the system size $N$, see Fig. \ref{fig:Fish}b. Similarly as for the negativity, the decay rate {\em decreases} with growing $m$. In fact by fitting to an exponential function $a\exp(-\gamma N)$, we obtain again a rate $\gamma$ that decays itself exponentially with $m$.


\textit{State Preparation.---}Having established that the C-GHZ state is robust against decoherence, we now discuss a possible experimental realization of these states using trapped ions. We consider the set-up of \cite{IonsBlatt}. The dominant source of noise in this case is correlated phase noise, which can be countered by using a decoherence free subspace. This can easily be combined with our approach by using C-GHZ states with logical qubits $|\tilde{GHZ}^{\pm}_m \rangle=\frac{1}{\sqrt{2}}(|01\rangle^{\otimes m/2} \pm |10\rangle^{\otimes m/2})$ for even $m$, thereby offering additional robustness against other noise processes as discussed above.

Entangling operations on a linear string of ions are performed using the multipartite M\o lmer-S\o rensen (MS) gate \cite{MS}, $U_n(\xi) = \prod_{k=1}^{n-1}\prod_{l=k+1}^{n} U_{kl}(\xi)$ with $U_{kl}(\xi)=e^{i \xi \sigma_x^k\otimes \sigma_x^l}$. Up to local unitaries, $U_n(\frac{\pi}{4})$ transforms the state $\ket{0}^{\otimes n}$ into a $n$-qubit GHZ state.

We now show how $U_n$ can be decomposed with single-qubit unitary operations \cite{MS} to obtain a C-GHZ state from the $n$--qubit GHZ state with $n=mN$. As a basic tool, we use that additional single qubit operations on a subset of ions, $Z(G) = e^{i \frac{\pi}{2} \sum_{k\in G}\sigma_z^k}$, allow one to reverse some of the two-qubit interactions induced by $U_n$, $Z(\{j\})U_n(\xi)Z^\dagger(\{j\}) = \prod_{\substack{k<l\\ k,l \neq j}} U_{kl}(\xi) \prod_{\substack{k<l\\ k\, \mathrm{or}\, l =j}}U_{kl}(-\xi)$. In particular, applying $Z(G)U_n(\frac{\xi}{2})Z^\dagger(G)$ with $G=\{1,2\ldots,\frac{n}{2}\}$ reverses all interactions between the two groups $\{1,\ldots,\frac{n}{2}\}$ and $\{\frac{n}{2}+1,\ldots,n\}$, but leaves the interactions invariant within the groups. An additional application of $U_n(\frac{\xi}{2})$ results in an effective phase $\xi$ within a group, and cancelation of all phases between the groups. One can now apply this method inductively, by splitting again each group in two halves and applying $Z(G')$ on $G'=\{1,\ldots\frac{n}{4}\} \cup \{\frac{3n}{4}+1,\ldots n\}$ while using MS-gates with $U_n(\frac{\xi}{4})$. In total, $N$ applications of $U_{Nm}\left(\frac{\pi}{4N}\right)$ and $N-1$ intermediate local rotations on groups of $Nm/2$ ions allow one to obtain an operation $V$ acting on $N$ blocks of size $m$, where no interactions takes place between blocks and a phase gate with phase $\frac{\pi}{4}$ acts between each pair of particles within each of the blocks. 
One can then generate first the GHZ state in the $z$-basis by applying $U_n(\frac{\pi}{4})$, and from there the C-GHZ state by applying $V$. This offers a scalable way of generating robust superpositions states in this ion-trap set-up.

Notice that the protocol is optimal --up to at most a factor of two-- with respect the the total time the entangling MS-gate has to be applied, which is $\frac{\pi}{2}$ in our scheme. The protocol is also efficient with respect to the required number of intermediate local unitary operations and MS-gates, which only scales linearly with $N$. In fact, for small $N$ we can show optimality with respect to the number of required MS-gates for protocols of this kind, consisting of MS-gates with arbitrary $\xi$'s and local $Z$-gates. We remark that more efficient preparation schemes are possible if MS-gates acting on subsets of qubits are available. In particular, a scheme involving a total of $O(m^2N)$ commuting two-qubit gates with interaction time $\frac{\pi}{4}$ can be found.



\textit{Discussion and Conclusion.---}
After illustrating the stability of the C-GHZ state with respect to various properties, one might wonder whether this is a generic feature of such encoded GHZ states, $\ket{0_L}^{\otimes N}+\ket{1_L}^{\otimes N}$. As long as $\ket{0_L}$ and $\ket{1_L}$ are orthogonal, the whole state keeps its ``catness''. However, we found for $m=3$ and $m=4$ that out of $10^6$ randomly chosen pairs of orthogonal quantum states (uniformly distributed with respect to the Haar measure), none showed a robustness comparable to the C-GHZ state with respect to decay of coherences ${\cal E}\ketbra{0_L}{1_L}^{\otimes N}$. Similarly, considering blocks of size $m$ for a GHZ state of size $mN$ also leads to an (exponentially) smaller robustness \cite{distill}. Perhaps more surprisingly, also using codewords of error correcting codes (a CSS code with $m=5$) for $|0_L\rangle, |1_L\rangle$, we found that {\em without} active error correction, the coherences still decay much faster as for the C-GHZ state \cite{note3}.

To summarize, we have introduced a class of quantum states that exhibit quantum properties similar to the GHZ state, but that are more stable under uncorrelated decoherence processes that we modeled by white noise. Although not explicitly shown here, equivalent robustness can be found for other decoherence models \cite{Fr11}. We have demonstrated that the trace norm of coherences, the lifetime of multipartite distillable entanglement, the decay of negativity and the Fisher information for quasi-local generators can be exponentially stabilized by considering blocks of a small number of qubits $m$ which are themselves prepared in a GHZ state. On the one hand, this may open new possibilities for practical applications of these states, e.g. in quantum communication or quantum metrology, as they offer an increased stability against noise and decoherence without need for active error correction. On the other hand, the existence of these states and their stability properties hints at the possibility to observe counter-intuitive quantum mechanical effects such as quantum superpositions of distinct states also at a macroscopic scale.

\textit{Acknowledgements.---} We thank L. Aolita, H.J. Briegel, O. G\"uhne, L. Maccone and M. Van den Nest for valuable discussions. This work was supported by the FWF and the European Union (QICS, SCALA, NAMEQUAM).


\begin{thebibliography}{99}
\bibitem{note1} One can indeed transform any single qubit noise process to single-qubit white noise by allowing for local unitary operations before and after the evolution, i.e. by adding additional noise \cite{Du05}. Hence stability under local white noise guarantees stability under any local noise in this sense.
\bibitem{note3}For $m=3$, the C-GHZ state is equivalent to a logical GHZ of size $\frac{N}{3}$ using the Shor code with blocks of 9 qubits. This is however not the setting we consider here.
\end{thebibliography}
\end{document}